# A strange feature of Bohm's theory of quantum motion

**T. Ouisse**

*Laboratoire de Spectrométrie Physique, Université Joseph Fourier Grenoble 1 and CNRS (UMR C5588), 140, rue de la physique, BP 87, 38042, Saint-Martin d'Hères Cedex, France*

November, 2002

**Abstract.** -   In the Bohm picture and for a *one-dimensional* analysis, we show that given an adequately chosen potential for characterising obstacles, one can derive laws of motion *formally* identical to that of special relativity. In such a hypothetical scheme, superluminal velocities are not forbidden, but a particle cannot collide with an obstacle with an average, superluminal velocity.

# 1. Introduction

In the pilot wave formulation of nonrelativistic quantum mechanics proposed by Bohm [1,2], one gives up the idea that the wave and particle notions are complementary to develop a picture in which both the point particles and the quantum wave coexist, the former being acted on by the latter through a quantum potential. A similar concept was first proposed in 1927 by De Broglie [3], but in Bohm's papers the theory is carried to a level at which it can reproduce all the empirical predictions of quantum mechanics [1,2,4,5]. In the Bohm's picture, one regains full determinism, and the hidden variables are not esoteric quantities, but simply the space coordinates of the particles. This is paid at the price of giving up locality, since the quantum field reacts instantaneously, and everywhere, to a change in the wave function [4,5]. Although the predictions of Bohmian mechanics and that of "standard" quantum theory are the same, the former approach has not received much attention so far, and is even often viewed by the scientific community as a superfluous and somewhat ideological interpretation of the quantum rules, since it does not seem to offer any new predictable result with respect to the conventional theory. Yet Holland has clearly identified why Bohm's proposal was a theoretical breakthrough: "it refutes the view that the individual facts of experience are *in principle* unimaginable" [5]. This was a credo of the Copenhagen interpretation, and it is not widely appreciated that the Bohm picture, whether physically correct or not, formally demonstrates that this credo is false. This point alone should justify a more thorough comparison between the implications of both interpretations. Reflections about hidden variable theories have indeed led Bell to formulate his celebrated theorem [6].

The purpose of this article is to address the problem of single particle wave interference in the Bohm's picture. First we derive a simple formula which gives the average velocity $v_{av}$ of a particle as a function of the quantum reflection coefficient characterising an obstacle. Then, we explicitly construct a reflection coefficient for characterising obstacles such that any massive particle will obey the laws of special relativity, given that the experimentally accessible velocity is the average one. In such a scheme, superluminal velocities are not forbidden in principle, but particles are forbidden by the quantum field to get close to obstacles at an average velocity greater than a given limit $c$. By construction the apparent mass and energy of the particles strictly follow the laws of special relativity. We stress that our analysis is restricted to one dimension (and indeed does not work in three). But through this mathematical example, we intend to demonstrate that a theory which is local in essence, and besides is non-Galilean, can be mimicked by a theory whose main feature is precisely non-locality, and which operates in a strictly Galilean scheme.

## 2. A brief summary of Bohm's theory

For the sake of clarity in this section we very briefly remind the central assumptions of Bohmian mechanics. The reader already aware of this interpretation of the quantum-mechanical rules can skip to section III.

In the Bohmian picture and single particle case, the wave function is written in the form $\psi=R\exp[iS/\hbar]$, so that Schrödinger's equation can be cast into the following two equations, the first one being analogous to the classical Hamilton-Jacobi equation of motion [1]:

$$\frac{\partial S}{\partial t}+\frac{(\nabla S)^2}{2m_0}+V+Q=0 \tag{1}$$

and

$$\frac{\partial P}{\partial t}+\nabla\left(P\frac{\nabla S}{m_0}\right)=0. \tag{2}$$

where $P=R^2=\psi\psi^*$, $V$ is the classical potential and $Q$ is the quantum potential [1]:

$$Q=-\frac{\hbar^2}{2m_0}\frac{\nabla^2 R}{R} \tag{3}$$

$P$ still represents a probability density in a statistical ensemble, and eq.(2) simply expresses probability conservation. It is postulated that the particle momentum $p$ is at all times given by

$$p=\nabla S, \tag{4}$$

so that eq.(1) also represents the equation of motion of the particle [1]. Both the $\psi$-field and the point particle are "real". What makes the particle quantum is that it is acted on not only by the potential *V*, but also by the $\psi$ wave through the quantum potential *Q*, according to eq.(1). From eq.(3), the particle motion does not depend on the magnitude of the wave function, but on its shape [5]. From (1) the equation of motion of the particle is [1]

$$m_0 \frac{d^2 \boldsymbol{x}}{dt^2} = -\nabla(V+Q) \qquad (5)$$

In practice, to calculate any particular particle trajectory, one first solves Schrödinger's equation to get $\psi$, and then differentiate *S* with respect to the space coordinates to get $\boldsymbol{p}$. The velocity is eventually integrated versus time, taking into account the proper initial conditions, so as to obtain the variation of position with time [5-10]. Bohm's theory can be generalised to an arbitrary number of particles [5]. It reproduces all the features of quantum mechanics but does not require observers to explain a measurement process [5]. What we do not know in a given experiment is the initial position of the detected particle, which even in a two-slit experiment follows a well-defined path (indeed a quite complicated one due to the action of the quantum potential [8]). Obviously, to reproduce the quantum-mechanical predictions the trajectories must exhibit some weird features (from a classical point of view). In the following we shall try to illustrate through a particular analytical example the relative perversity with which the Bohmian particles get close to obstacles, suddenly accelerating up to velocities values far above their average one, or simply refusing to move forward if they feel that they cannot pass.

## 3. Plane wave interference and construction of a special potential

We shall first be interested in solving the one-dimensional tunnelling case of a particle coming from minus infinity and crossing an energy barrier. For convenience the barrier is located from *x=0*. Take an incident, monochromatic beam of energy $E=\hbar k^2/2m_0=\hbar\omega$. For *x<0* the static wavefunction can be cast into the form

$$\psi(x) = \exp(ikx) + u\exp(-ikx) \tag{6}$$

where $u = \rho\exp(i\Phi)$ is a complex number of modulus $\rho$ and phase $\Phi$ ($\rho^2$ is the reflection coefficient). Introduction of the form (6) into eqs.(3,4) gives, after some calculation, the particle velocity $v$ and the quantum potential $Q$ for $x<0$:

$$v(x) = \frac{\hbar k}{m_0} \frac{1-\rho^2}{1+\rho^2+2\rho\cos(2kx-\Phi)} \tag{7}$$

$$Q(x) = \frac{\hbar^2 k^2}{2m_0}\left(1 - \frac{(1-\rho^2)^2}{(1+\rho^2+2\rho\cos(2kx-\Phi))^2}\right) = \frac{1}{2}m_0\left(v_0^2 - v(x)^2\right) \tag{8}$$

where $v_0 = \hbar k/m_0$. To illustrate how the particle passes through the barrier, one can consult, *e.g.*, the letter by Spiller *et al.* [9] for a simple analytical case (square barrier). In fact the particle does not have to tunnel through the energy barrier, for the quantum potential lowers it so as to permit the real particle to overcome a classically inaccessible region. Before reaching the obstacle, the interference between the incoming and reflected waves makes the quantum potential to oscillate (see Eq.(7,8)), so that the particle periodically accelerates and decelerates under the action of the $\psi$-field, and the velocity oscillates before reaching the barrier (such a feature is observable in the articles by Spiller et al. [9] and by Dewdney and Hiley [10]). If we calculate the maximum and minimum velocities, we find from eq.(7)

$$v_{\min} = v_0 \frac{1-\rho}{1+\rho} \tag{9}$$

and

$$v_{\max} = v_0 \frac{1+\rho}{1-\rho}. \tag{10}$$

Eq.(10) was first calculated by Leavens and Sala Mayato [7]. An increase in $\rho$ leads to a correlated increase in maximum velocity. As a side note to this letter, we stress that $v_{max}$ exceeds the light velocity $c$ if the reflection coefficient becomes greater than a simple combination of $v_0$ and $c$:

$$\rho \geq \frac{c-v_0}{c+v_0} \qquad (11)$$

Such a strange possibility was pointed out quite recently [7]. Keeping $v_0$ in a non-relativistic range, it is straightforward to adjust the tunnelling parameters so as to fulfil condition (11). Then, the interference between the incoming and reflected waves generates quantum potential wells deep enough for accelerating the particle up to a superluminal velocity, whereas from the side of "orthodox" quantum mechanics, embarrassing relativistic velocities do not show up anywhere, since any velocity expectation value remains largely sub-luminal. As a matter of fact, for a completely repulsive barrier $v_{max}$ becomes infinite [7]! But in fact the velocity field is not the only strange feature of the case depicted above. If one is now interested in finding the average velocity of the incident particle, one can straightforwardly integrate Eq.(7) to get the trajectory. Then, integrating over one period $\Delta L=\pi/k$ of the velocity field immediately gives the time $\Delta t$ required by the particle to travel over $\Delta L$, and the ratio $v_{av}=\Delta L/\Delta t$, which is independent of $x$ and thus is also the average velocity, is given by

$$v_{av} = v_0 \frac{1-\rho^2}{1+\rho^2} = v_0 \frac{T^2}{2-T^2} \qquad (12)$$

where $T^2$ is the transmission coefficient. It is then obvious that if the reflection coefficient tends to $1$, the particle gets immobilised, no matter how far it lies from the obstacle. In other words, when the particle cannot pass, it does not come! And if the transmission coefficient is small the average velocity is proportional to it. Here we note that in general $\rho$ depends on $v_0$, and it is let to the reader to find in some simple situations how curious can become the relationship between $v_0$, which in fact is *not* the incident velocity, but a quantity which characterises the incident energy ($E=m_0v_0^2/2$), and the real, average velocity of the particle (for instance, in the case of a thick and square tunnelling barrier, the average velocity gets proportional to the cube of $v_0$!).

Now suppose that one lives in a strange one-dimensional quantum world, where all physical obstacles are characterised by an extremely sharp transmission resonance, say $T^2=1$ at an energy $E_T=m_0c_0^2/2$ where $c_0$ is an arbitrary constant. Then the only particles which could enter in physical contact with those obstacles would be the ones with an average, relative velocity precisely equal to $c_0$… In such a case the energy of the particles which can collide can take no other value than $E_T$, but in the following we shall demonstrate that one can construct two simple reflection coefficients, from which one can recovers the whole set of the laws of special relativity for massive particles and photons. We make the assumption that the physical quantities which can be known by the observer are the average velocity (which he can measure using synchronised clocks localised respectively at the particle emitter and at the obstacle with which the particle interacts) and the real particle energy (which he can measure by energy transfer to the obstacle in any adequate experiment). Besides, we stress again that as in the calculations above, we assume that the relevant solution to the Schrödinger's equation is the static one, *i.e.* there is no wave packet (this means that the particle never "freely" propagates and its movement is influenced by the obstacle from the start [5]). Then, we characterise any obstacle to massive particles by a reflection coefficient $\rho^2$ given by

$$\rho^2 = \frac{v_0^3 - c\sqrt{v_0^4 - 4c^4}}{v_0^3 + c\sqrt{v_0^4 - 4c^4}} \qquad (13)$$

(here $c$ is the usual velocity limit of special relativity). The form (13) imposes that $v_0$ cannot be smaller than $c\sqrt{2}$, but this precisely corresponds to the rest energy of special relativity $E_0 = m_0 c^2$. The corresponding transmission coefficient $T^2 = 1 - \rho^2$ is plotted in Fig.1 as a function of energy. It exhibits a resonance slightly above the rest energy (for $E = \sqrt{3} E_0$). From (13) and (12) it is straightforward to derive that any of the particles which come into contact with the obstacle must obey the following features:

(1) The average particle velocity $v_{av}$ (as given by (12) and (13)) of the particles which can collide with the obstacle cannot exceed $c$.

(2) The energy of the particles which can collide with the obstacle is

$$E = mc^2 \qquad (14)$$

with

$$m = \frac{m_0}{\sqrt{1 - \frac{v_{av}^2}{c^2}}} \,. \qquad (15)$$

For photon-like particles we take the reflection coefficient

$$\rho^2 = \frac{v_0 - c}{v_0 + c} \qquad (15)$$

Then it is straightforward to check from Eq.(12) that those photon-like particles can only travel towards obstacles at a constant average speed, precisely equal to *c*, and independent of their energy. It is worth noticing that Eq(15) is nothing but the high $v_0$ limit of eq.(13), so that provided that particles have a very small mass, they become photon-like even for small energy values.

The features above are nothing but the usual laws of special relativity. Besides, the reader can easily check that Eq.(13) is indeed the only form that can be taken by the reflection coefficient if Eqs.(14,15) are to be verified. It is thus remarkable that any of the values given by Eq.(13) remains lower than 1, as expected for a physical reflection coefficient, because this condition was not imposed at any stage of our calculations. What must be understood here is that the velocity limit is not a constraint imposed by a non-Galilean space-time configuration. It is imposed by the dynamical interaction between the particle, the quantum field and the obstacle (whose potential must correspond to a reflection coefficient given by eq.(13)). It must be stressed that the whole picture is deeply non-local: one obtains the results above if the quantum field propagates and reacts instantaneously. Besides, one cannot speak of a particle moving alone, but one must consider *from the beginning* the interaction it may have with its surroundings, because the surrounding environment instantaneously determinates the quantum field which in turn governs the particle movement.. It is quite disturbing (at least for the author of this letter) that the theory of special relativity, which on the one hand is local in essence, and on the other hand implies a specific space-time configuration, can be mimicked by a theory which is Galilean and non-local. And most of all, this theory is not constructed *ab initio* for this purpose (except, of course, in our choice of the reflection coefficient), but is merely an interpretation of an existing theory, quantum mechanics, whose experimental success remains undisputed since its birth. The calculations above show that it is possible to turn the geometrical constraints of Einstein-Minkowski space-time into dynamical ones, imposed upon particle motion by a specific potential. We hope that our hypothetical example has at least the merit to demonstrate that drastically opposite physical assumptions can lead to the same formal predictions. Obviously the first point to be examined with further detail is the form in real space of the potential defined by the reflection coefficient (13). Does it correspond to a finite-range potential, or can at least be approximated on a large incident energy range by the reflection coefficient of a finite range potential ? There is no doubt in the author's mind that this can be achieved at least on a restricted energy range (what may be a source of problem is the fact that the reflection coefficient tends to 1 when the energy goes to infinity). To express our special potential in real space one has to use the tools available from inverse problem theory, such as, *e.g.*, to solve the Gel'fand Levitan equation [11] (we note that the reflection coefficient (13) can be easily expanded into a Laurent series of *(v-v_0)*). This point is let for future investigation. It is worth noticing that we have derived this result in one dimension. Unfortunately, the same reasoning cannot be applied to a central potential in a three-dimensional world. Extending our calculation to an obstacle characterised by a central potential is equivalent to replace $\rho_2$ by $|f_k(\theta=\pi,\varphi)|^2/r^2$ in our previous formulae, where $r$ is the radial distance from the obstacle and $|f_k(\theta=\pi,\varphi)|^2$ is the differential cross section at angle $\theta=\pi$. Unless the differential cross section is singular for backscattering and zero for all other angles, the term in *$1/r^2$* makes the influence of the backscattered wave on the incident average velocity to vanish at large distances, so that it is not possible to obtain constant, average velocities in that case.

## 4. Conclusion

We have given an analytical form of a reflection coefficient which provides the ability to *formally* recover the basic features of special relativity from Bohm's theory of quantum motion, even though the latter theory is non-local and operates in Galilean space-time. The geometrical constraints of relativistic space-time are turned into dynamical constraints, imposed by a special potential acting upon the particle motion through the pilot wave. We stress that our mathematical example and results are restricted to a one-dimensional analysis. In any case, that the basic equations of a local theory operating in Einstein-Minkowski space-time can be mimicked by a non-local, Galilean theory is a mathematical and physical curiosity.

# Figure captions

**Fig.1:** Variation versus energy of the transmission coefficient $T^2$ which allows one to obtain from Bohm's Theory laws of motion formally identical to that of special relativity.

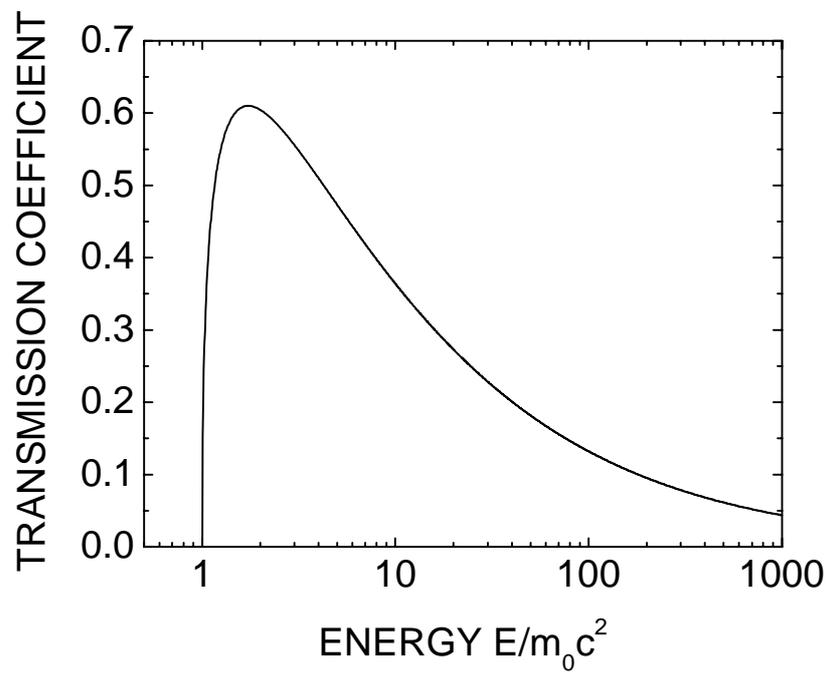